\documentclass[letterpaper,twocolumn,10pt]{article}
\usepackage{usenix2019,epsfig,endnotes}

\usepackage{booktabs}

\usepackage{marvosym}

\usepackage{array}

\usepackage{graphicx}
\usepackage{epstopdf}
\usepackage{subfig}
\usepackage{comment}

\usepackage{amsmath}
\usepackage{arydshln}
\usepackage{algorithm}
\usepackage{algorithmic}
\renewcommand{\algorithmiccomment}[1]{\bgroup\hfill//~#1\egroup}

\usepackage{url}

\usepackage{color}
\definecolor{MyGreen}{RGB}{0,100,0}
\definecolor{MyRed}{RGB}{255,0,102}

\usepackage[colorlinks=true, linkcolor=red, anchorcolor=red, citecolor=red]{hyperref}

\usepackage{lipsum}
\usepackage{titlesec}
\usepackage{makecell}

\titlespacing\section{0pt}{7pt plus 4pt minus 2pt}{5pt plus 2pt minus 2pt}
\titlespacing\subsection{0pt}{7pt plus 3pt minus 2pt}{5pt plus 2pt minus 2pt}
\titlespacing\subsubsection{0pt}{7pt plus 3pt minus 2pt}{6pt plus 2pt minus 2pt}

\newcommand{\labcolor}[1]{\textcolor{red}{#1}}
\newcommand{\final}[1]{\textcolor{black}{#1}}
\newcommand{\atc}[1]{\textcolor{black}{#1}}
\newcommand{\reencryption}[1]{\textcolor{black}{#1}}

\begin{document}

\hyphenpenalty = 8000
\tolerance = 2000

\date{}

\title{\Large \bf A Secure and Persistent Memory System for Non-volatile Memory~\vspace{-0.4cm}}

\author{
{\rm Pengfei Zuo{\textsuperscript{* \textdagger}}, Yu Hua{\textsuperscript{*}}, Yuan Xie{\textsuperscript{\textdagger}}
}\\
{\textsuperscript{*}}Huazhong University of Science and Technology \\
{\textsuperscript{\textdagger}}University of California, Santa Barbara    \\
}

\maketitle


\subsection*{Abstract}
\noindent
In the non-volatile memory, ensuring the security and correctness of persistent data is fundamental. However, the security and persistence issues are usually studied independently in existing work. To achieve both data security and persistence, simply combining existing persistence schemes with memory encryption is inefficient due to crash inconsistency and significant performance degradation. To bridge the gap between security and persistence, this paper proposes \textbf{SecPM}, a \textbf{Sec}ure and \textbf{P}ersistent \textbf{M}emory system, which consists of a counter cache write-through (CWT) scheme and a locality-aware counter write reduction (CWR) scheme. Specifically, SecPM leverages the CWT scheme to guarantee the crash consistency via ensuring both the data and its counter are durable before the data flush completes, and leverages the CWR scheme to improve the system performance via exploiting the spatial locality of counter storage, log and data writes. We have implemented SecPM in gem5 with NVMain and evaluated it using five widely-used workloads. Extensive experimental results demonstrate that SecPM reduces up to half of write requests and speeds up the transaction execution by $1.3\times \sim2.0 \times$ via using the CWR scheme, \atc{and achieves the performance close to an un-encrypted persistent memory system for large transactions.}

\section{Introduction}
\noindent
As DRAM suffers from limited scalability and high power leakage~\cite{mueller2005challenges,thoziyoor2008comprehensive}, non-volatile memories (NVM), such as PCM~\cite{wong2010phase}, ReRAM~\cite{akinaga2010resistive}, STT-RAM~\cite{apalkov2013spin}, and 3D XPoint~\cite{intel3dxpoint}, become promising candidates of the next-generation main memory. NVM has the advantages of high scalability, high density, and near-zero standby power. However, two fundamental issues need to be addressed in order to effectively use NVM in memory systems, i.e., data persistence and security.

\emph{\textbf{$\bullet$ Persistence Issue.}} The non-volatility of NVM enables data to be persistently stored into main memory for instantaneous failure recovery. In order to ensure the correctness of persistent data, crash consistency guarantee is fundamental~\cite{pelley2014memory,volos2011mnemosyne}, which needs to achieve the correct recovery of persistent data in case of a system failure, e.g., power failure and system crash. Specifically, NVM systems typically contain volatile storage components, e.g., CPU caches and possible DRAM. If a system failure occurs when a data structure in NVM is being updated, the data structure may be left in a corrupted state. Moreover, modern processor and memory controller usually reorder memory writes. The partial update and reordering cause the crash inconsistency in NVM~\cite{leewort2017,yang2015nv}. Hence, cache line flush, memory barrier, and log-based mechanisms are used to ensure the crash consistency~\cite{liu2017dudetm,xu2016nova}.

\emph{\textbf{$\bullet$ Security Issue.}} The non-volatility of NVM also causes the security problem of data remanence vulnerability~\cite{awad2016silent,young2015deuce}, since NVM still retains data after systems are powered down. \final{In general, when using encryption to protect data security, the encrypted data are stored in disks}, while raw data are retained in main memory~\cite{AMD2016}. In the legacy DRAM main memory, if a DRAM DIMM is stolen, data are quickly lost due to the volatility. Unlike it, \final{if an NVM DIMM is stolen}, an attacker can directly stream out the data from the DIMM. Hence, memory encryption becomes important to ensure the data security in NVM. Counter mode encryption~\cite{Lipmaa2000CTR,young2015deuce} is usually used in secure NVM, due to the low decryption latency and high security level.

However, existing schemes addressing the persistence issue~\cite{kolli2016high,pelley2014memory,Seo2017FSP,volos2011mnemosyne,xu2016nova} usually fail to efficiently use memory encryption in NVM systems. On the other hand, existing schemes addressing the security issue~\cite{awad2016silent,chhabra2011nvmm,swami2016secret,young2015deuce,zuo2018dewrite} are inefficient to guarantee the data consistency. To achieve both data persistence and security in NVM, simply combining existing persistence schemes with memory encryption does not work due to the following challenges.

\emph{\textbf{$\bullet$ Consistency Challenge.}} In order to guarantee crash consistency, it is essential to use cache line flush and memory barrier instructions to persist data into NVM with correct ordering. In the counter mode encryption, each memory line is encrypted and decrypted using a counter, and the counter increases one on each write~\cite{Lipmaa2000CTR}. Thus each write in the secure NVM has to persist two data including the data itself and its counter~\cite{liucrash2018}. The data is evicted from the CPU caches and the counter is evicted from the counter cache managed by the memory controller. \atc{The two writes have to be simultaneously persisted to ensure that the persistent data can be correctly decrypted across system failures. For example, if the system fails after the encrypted data
is persisted into NVM, but before its counter has been persisted, the persisted data cannot be decrypted without the correct counter during the recovery from a system failure. However, current computer systems fail to simultaneously persist the two writes, thus resulting in an inconsistent state on a system failure~\cite{liucrash2018,swami2018acme}. This is because} cache line flush and memory barrier instructions only ensure the data from the CPU caches to be correctly persisted into NVM, which fail to operate the counter cache and hence cannot ensure the crash consistency of counters.

\emph{\textbf{$\bullet$ Performance Challenge.}} Each data write in the secure NVM generates two NVM write requests (one writes the data and the other writes its counter), which significantly degrades the system performance. Because writes to NVM usually incur much higher latency than reads (i.e., $3\sim 8 \times$)~\cite{qureshi2009enhancing,zhou2009durable}, and are in the critical path of application execution due to persistence requirements~\cite{volos2015quartz,volos2011mnemosyne}. Moreover, more write requests also significantly increase the latency of read requests to the same bank. When a write request is served by an NVM bank, the following read and write requests to the same bank are blocked and have to wait until the current write request is completed~\cite{qureshi2010improving}.

To bridge the gap between security and persistence, this paper proposes a secure and persistent memory system, called SecPM. SecPM proposes to use a simple yet effective counter cache write-through (CWT) scheme with a register to ensure the crash consistency of both data and counters. CWT appends the counter of the data in the write queue during encrypting the data, which ensures the counter is durable before the data flush completes. \atc{Thus the two writes for the data and its counter are performed in an atomic manner.} Moreover, SecPM leverages a locality-aware counter write reduction (CWR) scheme to improve the system performance. CWR explores and exploits the spatial locality of counter storage, log and data writes to merge write requests from counters in the write queue, thus significantly reducing the number of NVM write requests.
In summary, this paper makes the following contributions:
\begin{itemize}
  \item \textbf{\emph{Simple yet Effective Consistency Guarantee.}} \atc{Existing work adds new primitives in the programming language to explicitly flush programmer-specified counter cache lines in the counter cache for consistency guarantee~\cite{liucrash2018}, which however incurs modifications on both hardware and software layers.} The novelty of SecPM is to leverage a simple yet effective counter cache write-through (CWT) scheme with a register to guarantee the crash consistency of secure NVM. Our scheme is able to ensure both data and counters to be simultaneously persisted before a CPU cache line flush instruction is retired.
  \item \textbf{\emph{Significant Performance Improvement.}} SecPM leverages a locality-aware counter write reduction (CWR) scheme to improve system performance via significantly reducing the number of NVM write requests from counters. Extensive experimental results demonstrate that SecPM reduces up to half of write requests, and speeds up the transaction execution by $1.3\sim2.0$ times by using the CWR scheme, \atc{and achieves the performance close to an un-encrypted persistent memory system for large transactions.}
  \vspace{-5px}
  \item \textbf{\emph{Programmer-transparent Implementation.}}
  The implementation of SecPM only needs to perform slight modifications on the hardware layer without any modifications on the software layer, e.g., programming language and compiler, which are transparent for programmers and applications. Thus the programs and applications running on an un-encrypted NVM can be directly executed on a secure NVM with SecPM.
  \vspace{-5px}
\end{itemize}



\vspace{-5px}
\section{Background and Motivation}
\label{section2}
\vspace{-3px}
\begin{figure}[b]
  \vspace{-8px}
  \centering
    \includegraphics [width=0.33\textwidth]{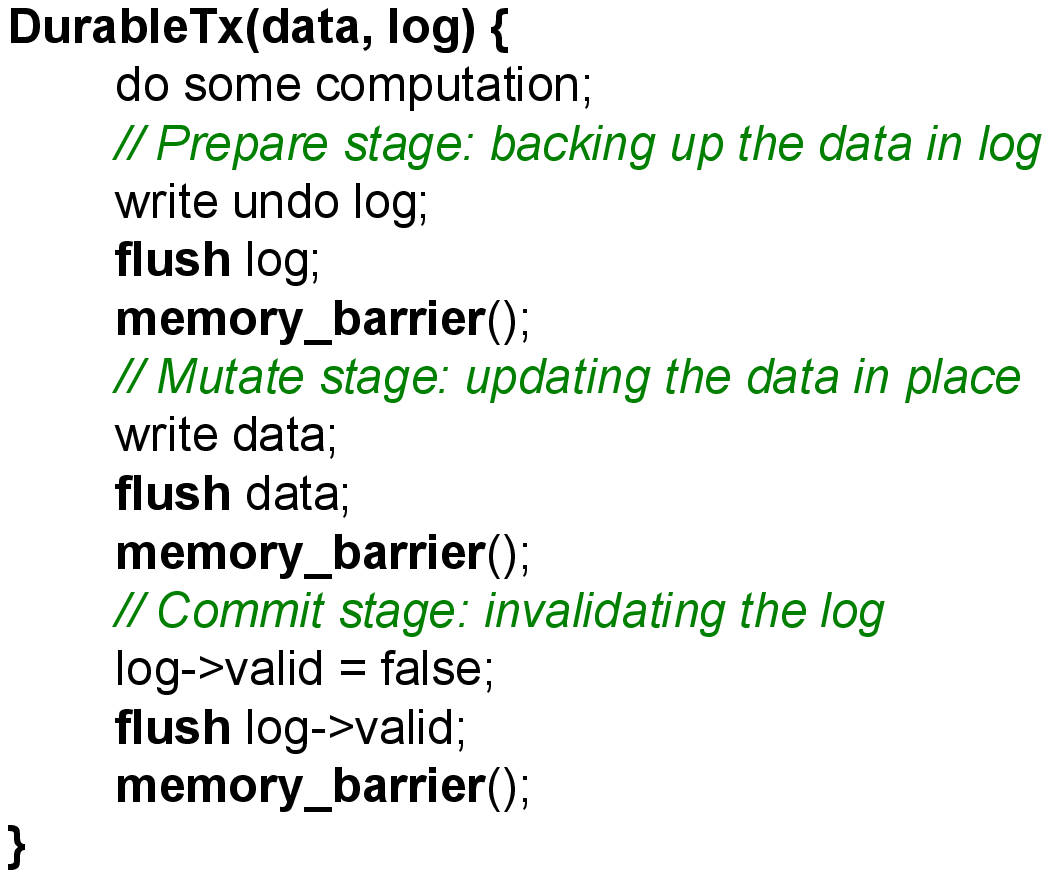}
    \vspace{-6px}
    \caption{\label{fig:transaction} Steps implementing a durable transaction.}
   \vspace{-2px}
\end{figure}


\subsection{Consistency Guarantee for Persistence}
\label{background:consistency}
\noindent
In order to correctly persist data into NVM, it is important to guarantee the crash consistency. Durable transaction~\cite{kolli2016high,liu2017dudetm} is a commonly used solution for crash consistency guarantee, which enables a group of memory updates to be performed in an atomic manner. Figure~\ref{fig:transaction} shows the \atc{pseudo-code that an application implements} a durable transaction with undo logging, which include three stages, i.e., prepare, mutate and commit. Specifically, the prepare stage creates a log entry to back up the data to be written; the mutate stage modifies the data in place; the commit stage invalidates the log entry created in the prepare stage. Whichever stage a system failure occurs, the application can be recoverable in a consistent state since at least one of the log and data are consistent.

As modern CPU and memory controller usually reorder memory writes, the durable transaction uses cache line flush and memory barrier instructions to enforce write ordering~\cite{kolli2016high,liu2017dudetm,pelley2014memory}. The flush instructions including \verb"clflush", \verb"clflushopt", and \verb"clwb" explicitly flush a dirty CPU cache line into the write queue of the memory controller. The memory barrier instructions including \verb"mfence" and \verb"sfence" order the memory operations via blocking the memory operations after the fence, until the memory operations before the fence complete. The \verb"pcommit" instruction was initially used to force the write requests in the write queue into NVM but was deprecated later by Intel~\cite{pcommit2016}, due to the use of asynchronous DRAM refresh (ADR) mechanism~\cite{inteladr2017,liucrash2018,shin2017proteus}. ADR is able to persist the write requests in the write queue into NVM in case of a system failure via the battery backup. Therefore, the cache lines reaching the write queue are considered durable.

\subsection{Memory Encryption for Security}
\noindent
\noindent
\atc{Since NVM suffers from the data remanence vulnerability,} memory encryption is non-trivial for NVM to ensure data security.

\subsubsection{Security Guarantee of Encryption}
\begin{figure}[h]
  \centering
    \includegraphics [width=0.46\textwidth]{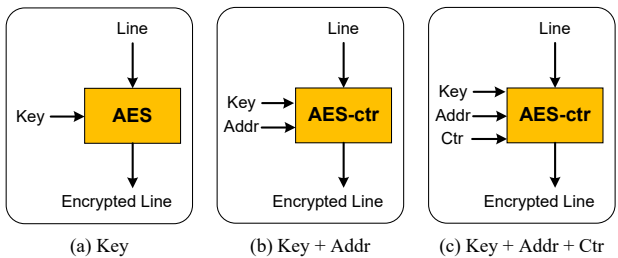}
    \caption{\label{fig:background-security-level} The encryption methods (a) using a global key; (b) using a key and line address; (c) using a key, line address and a counter (ctr).}
\end{figure}

\noindent
A straightforward method to encrypt a memory line is to use a block cipher
algorithm, e.g., AES~\cite{daemen2013design}, with a global key, as shown in Figure~\ref{fig:background-security-level}\labcolor{a}. However, an attacker can know which lines have the same content via simply comparing encrypted lines, since all lines are encrypted using the same key, which is vulnerable to dictionary and replay attacks~\cite{awad2016silent,young2015deuce}. Using the key along with the line address to encrypt each line is a more secure method, as shown in Figure~\ref{fig:background-security-level}\labcolor{b}, which can ensure that different lines are encrypted with different keys. However, this method is still vulnerable to the dictionary attack for a single line, if an attacker monitors consecutive writes to this line.

A secure method is to encrypt each line by using the key and line address in conjunction with a per-line counter, as shown in Figure~\ref{fig:background-security-level}\labcolor{c}. The counter increases by one on each write and hence consecutive writes to the same line are encrypted with different keys, achieving high security level.

\subsubsection{Latency Reduction via One Time Pad}
\begin{figure}[t]
\vspace{5px}
  \centering
    \includegraphics[width=0.45\textwidth]{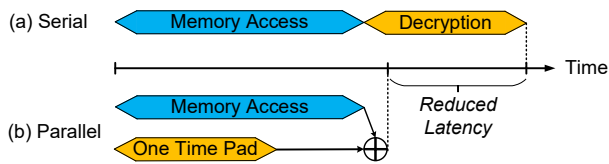}
    \caption{\label{fig:latency-counter-mode} Reducing the decryption latency using One Time Pad (OTP).}
    \vspace{-5px}
\end{figure}

\noindent
As memory reads are in the critical path of program execution, memory encryption causes the high decryption latency after each memory read due to serial execution, as shown in Figure~\ref{fig:latency-counter-mode}\labcolor{a}, thus significantly degrading the system performance. Counter mode encryption~\cite{Lipmaa2000CTR} is hence proposed to reduce the decryption latency from the critical path of memory reads via leveraging the One Time Pad (OTP) technique, and hence has been widely used in encrypted memory systems~\cite{awad2016silent,chhabra2011nvmm,swami2016secret,young2015deuce,zuo2018dewrite}. The main idea is to compute an OTP in parallel with a memory read, and then XOR the OTP with the ciphertext data to generate the plaintext, thus hiding the decryption latency in the memory access latency, as shown in Figure~\ref{fig:latency-counter-mode}\labcolor{b}.

\subsubsection{Operations of Counter Mode Encryption}
\begin{figure}[h]
  \centering
    \includegraphics[width=0.39\textwidth]{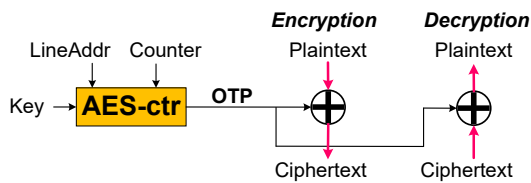}
    \caption{\label{fig:OTP} The encryption and decryption process in counter mode encrpytion.}
    \vspace{-5px}
\end{figure}

\noindent
The security of counter mode encryption is based on the premise that each OTP is never reused for data encryption~\cite{Lipmaa2000CTR,swami2016secret,young2015deuce,zuo2018dewrite}. To ensure this, the counter mode encryption uses a secret key, the line address and the per-line counter to generate the OTP through the AES circuit, as shown in Figure~\ref{fig:OTP}.
For a memory write, the cache line to be written is encrypted by XORing its content with the OTP. To read a memory line, we decrypt it by XORing its content with the OTP. All counters are retained in main memory.
In order to reduce the generation time of OTPs, the memory controller manages an on-chip counter cache to buffer the recently-accessed counters~\cite{liucrash2018,swami2016secret,zuo2018dewrite}.

 \vspace{-3px}
\subsection{The Gap between Persistence $\&$ Security}
\label{background:challenges}

\noindent
Each cache line flushed from CPU caches in the encrypted NVM produces two write requests: one for the data and the other for the counter. This characteristic not only degrades the system performance since NVM needs to deal with more write requests, but also incurs the crash inconsistency problem. The reason is that the data is evicted from the CPU caches but its counter is evicted from the counter cache. However, the cache line flush instruction only flushes the data in CPU caches, but fails to operate the counter cache. The memory barrier instruction \atc{only ensures the ordering of the CPU cache line flushes, but cannot work for the} counter writes~\cite{liucrash2018,swami2018acme}. As a result, \atc{the data and its counter cannot be simultaneously persisted,} and thus the persisted data cannot be decrypted without correct counters during the recovery from a system failure.

We analyze the recoverability when a system failure occurs in the different stages of a durable transaction executed in an encrypted NVM, as shown in Table~\ref{table-observation}. We observe when a system failure occurs in the mutate and commit stages, the data are unrecoverable. Specifically, when a system failure occurs in the prepare stage, the data contents and the counters encrypting the data are unmodified and correct, which are in a consistent state. However, when a system failure occurs in the mutate stage, the data are not updated completely and become wrong. The contents of the log are correct due to the use of log flush and memory barrier instructions, but whether the counters encrypting the log are correctly persisted is unknown, since the cache line flush and memory barrier instructions fail to operate the counters stored in the counter cache. Hence, during a recovery, the log cannot be decrypted due to no correct counters, thus failing to recover the logged data. For the same reasons, when a system failure occurs in the commit stage, the correctness of both log and data counters are unknown, and hence the data are unrecoverable.

\begin{table}[t]
\scriptsize
  \begin{center}
     \caption{\label{table-observation} The recoverability when a system failure occurs in the different stages of a transaction.}
    \vspace{-3px}
    \begin{tabular}{p{0.72cm}<{\centering}|p{0.85cm}<{\centering}|p{0.85cm}<{\centering}|p{0.85cm}<{\centering}|p{0.85cm}<{\centering}|c}
\hline
    \textbf{Stage} & \textbf{Log Content} & \textbf{Log Counter}   & \textbf{Data Content}  & \textbf{Data Counter} &   \textbf{Recoverable ?}          \\
\hline
    Prepare     & Wrong  & Wrong  & Correct & Correct & Yes     \\
    Mutate      & Correct & \textbf{Unknown}  & Wrong   & Wrong & \textbf{No}     \\
    Commit      & Correct & \textbf{Unknown} & Correct & \textbf{Unknown} & \textbf{No}    \\
\hline
    \end{tabular}
    \end{center}
    \vspace{-20px}
\end{table}

To address the inconsistency problem in the encrypted NVM, Liu et al.~\cite{liucrash2018} proposed the novel concept of the selective counter-atomicity \atc{for the first time}, which indicates that either both data and its associated counter have been simultaneously persisted or not. To efficiently implement the counter-atomicity, software and hardware layers need to be modified~\cite{liucrash2018}. First, two new primitives including \verb"CounterAtomic" variable and \verb"counter_cache_writeback()" function, are added in the programming language  to explicitly flush programmer-specified counter cache lines. Second, the compiler is modified to support the new primitives. Third, a counter write queue is added into the memory controller to schedule the counter and data writes. As a result, the programs initially running on a system with the un-encrypted NVM cannot directly run on a system with the secure NVM.
\atc{Our work aims for a different design goal. The proposed SecPM is a programmer-transparent solution} to guarantee the crash consistency of the secure NVM without the needs of modifications on programming language and compiler, which is detailed in the rest of the paper.

\section{The SecPM Design}
\label{section3}

\begin{figure}[t]
  \vspace{2px}
  \centering
    \includegraphics [width=0.27\textwidth]{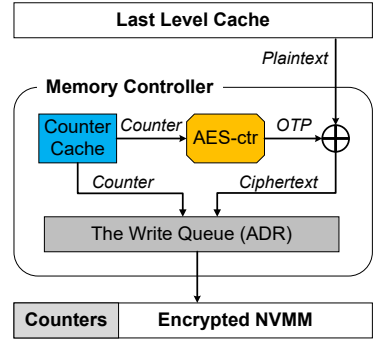}
    \caption{\label{fig:secpm-architecture} The hardware architecture of SecPM. \textit{(The asynchronous DRAM refresh (ADR) mechanism is able to persist write requests in the write queue into NVMM on a system failure via the battery backup. Therefore, the cache lines reaching the write queue are considered durable.)}}
    \vspace{-11px}
\end{figure}

\subsection{\atc{Threat Model}}
\label{threat-model}
\noindent
\atc{Our threat model is similar to existing work on secure NVM~\cite{awad2016silent,liucrash2018,swami2016secret,young2015deuce,zuo2018dewrite}, which aims to protect NVM from two well-known physical access based
attacks, including stolen DIMM and bus snooping attacks. In the stolen DIMM attack, since NVM still retains data after systems are powered down, an attacker can easily stream out the data stored in the NVM after stealing the NVM DIMM. In the bus snooping attack, since NVM is accessed through the memory bus, an attacker can insert a bus snooper to obtain the data through the bus. We do not consider bus tampering attacks in the threat model like existing work~\cite{awad2016silent,liucrash2018,swami2016secret,young2015deuce,zuo2018dewrite}. These attacks can be defended via Merkle Trees based authentication techniques~\cite{suh2003efficient,ye2018osiris}, which are orthogonal to our work.}

\subsection{\atc{An Architectural Overview}}
\noindent
\atc{To bridge the gap between security and persistence in NVM}, we propose SecPM, a \textbf{Sec}ure \textbf{P}ersistent \textbf{M}emory system, which leverages a simple yet effective counter cache write-through scheme ($\S$\ref{design:write-through}) with a register to guarantee the crash consistency of both data and its counter, and a counter write reduction scheme ($\S$\ref{design:write-reduction}) to significantly reduce the number of write requests for improving the system performance. \atc{Moreover, SecPM guarantees the consistency of page re-encryption that handles counter overflow by reusing the ADR mechanism ($\S$\ref{sec:page-re-encryption}).}


The hardware architecture of SecPM is shown in Figure~\ref{fig:secpm-architecture}. When CPU issues a flush instruction, the corresponding cache line is evicted from the last level cache to the memory controller. The memory controller encrypts the cache line using a counter and then appends the encrypted cache line in the write queue. The proposed counter cache write-through (CWT) scheme is performed in the counter cache, and the counter write reduction (CWR) scheme is performed in the write queue. As a whole, SecPM only performs slight hardware modifications on the memory controller, which are transparent for programmers and applications. Thus programs and applications running on an un-encrypted NVM can be directly executed on a secure NVM with SecPM.

\subsection{Crash Consistency Guarantee}
\label{design:write-through}
\subsubsection{Counter Cache Write Through Scheme}
\noindent
SecPM employs a counter cache write-through (CWT) scheme in the counter cache, which writes each dirty counter in the counter cache, and simultaneously writes the counter copy in the write queue. We further show how to schedule the cache line flush via using the simple CWT scheme to ensure the crash consistency of the secure NVM.

\begin{figure}[t]
  \centering
    \includegraphics [width=0.48\textwidth]{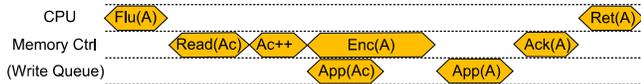}
    \caption{\label{fig:sequence} The sequence that the memory controller deals with a cache line flush by using the counter cache write-through scheme. \textit{(\textbf{Flu(A)}: flushing the cache line \textbf{A} into NVM; \textbf{Ac}: the counter of \textbf{A}; \textbf{App(Ac)}: appending \textbf{Ac} in the write queue; \textbf{Enc(A)}: encrypting \textbf{A}; \textbf{Ret(A)}: the flush of \textbf{A} is retired.)}}
    \vspace{-9px}
\end{figure}

Figure~\ref{fig:sequence} shows the sequence diagram that the memory controller deals with a cache line flush in SecPM. When the CPU issues a flush for cache line $A$ (\emph{Flu(A)}), the memory controller reads the counter of $A$ from the counter cache (\emph{Read(Ac)}), and then adds the counter by one (\emph{Ac++}). The updated counter is used to encrypt $A$ (\emph{Enc(A)}). During the encryption, the updated counter is written back to the counter cache, and simultaneously appended in the write queue (\emph{App(Ac)}) via the \final{CWT} scheme.
After the encrypted $A$ is appended in the write queue (\emph{App(A)}), the memory controller sends an ack (\emph{Ack(A)}) to the CPU, and the cache line flush is retired (\emph{Ret(A)}). From Figure~\ref{fig:sequence}, we observe that the counter encrypting a CPU cache line has been already appended in the write queue before the cache line flush completes via the CWT scheme. Hence, if the contents of log and data have been persisted and become correct, their counters have also been persisted.

\begin{table}[h]
\scriptsize
  \begin{center}
     \caption{\label{table-design} The recoverability when a system failure occurs in the different stages of a transaction in SecPM.}
    \begin{tabular}{p{0.72cm}<{\centering}|p{0.85cm}<{\centering}|p{0.85cm}<{\centering}|p{0.85cm}<{\centering}|p{0.85cm}<{\centering}|c}
\hline
    \textbf{Stage} & \textbf{Log Content} & \textbf{Log Counter}   & \textbf{Data Content}  & \textbf{Data Counter} &   \textbf{Recoverable ?}          \\
\hline
    Prepare     & Wrong  & Wrong  & Correct & Correct & Yes     \\
    Mutate      & Correct & Correct  & Wrong   & Wrong & Yes     \\
    Commit      & Correct & Correct & Correct & Correct & Yes    \\
\hline
    \end{tabular}
    \end{center}
   \vspace{-8px}
\end{table}

Table~\ref{table-design} shows the recoverability when a system failure occurs in the different stages of executing a durable transaction in SecPM. In the prepare stage, the contents and counters of data are un-modified and thus correct. When the prepare stage completes, it means that all contents of the log have been successfully flushed and persisted. As we demonstrate above, if the contents of log have been persisted, their counters have also been persisted by using the CWT scheme. Hence, both the contents and counters of the log have been persisted and thus are correct in the mutate stage.
For the same reason, in the commit stage, the contents and counters of the log and data are correct.
We observe that at least one of log and data are correct whichever stage a system failure occurs in, and hence the system can be recoverable in a consistent state.

\emph{\textbf{Recovery:}} \atc{After a system failure occurs, the recovery scheme of transactions in SecPM is the same as that in an un-encrypted persistent memory.} During the recovery, \atc{the application} scans the log region, and checks each log entry. For each log entry, \atc{the application} checks whether the log is complete via the log-end tag of a transaction. If the log is incomplete, it means the system crashes in the prepare stage of the transaction, in which the original data are correct. \atc{The application directly abandons} the incomplete log. If the log is complete, it means the system crashes in the mutate or commit stage, in which the log is correct. \atc{The application} undoes the data via the log. Therefore, \atc{the application} can be recovered in a consistent state in SecPM.

\begin{figure}[t]
  \centering
    \includegraphics [width=0.48\textwidth]{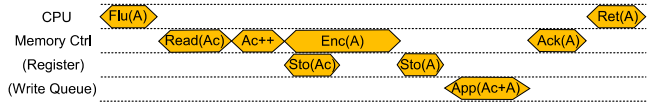}
    \caption{\label{fig:sequence-with-register} The sequence that the memory controller deals with a cache line flush by using the counter cache write-through scheme with a register. \textit{(\textbf{Sto(Ac)}: storing \textbf{Ac} in the register.)}}
    \vspace{-6px}
\end{figure}
\subsubsection{A Register for Atomic Write Consistency}
\noindent
In addition to using durable transaction, some existing work, e.g., wB$^+$-tree~\cite{chen2015persistent}, WORT~\cite{leewort2017}, FAST$\&$FAIR~\cite{hwang2018endurable}, and level hashing~\cite{222591}, exploits the atomic write of NVM to ensure the crash consistency of data structures, thus avoiding the overhead of logging or copy-on-write mechanisms. However, simply performing the proposed counter cache write-through (CWT) scheme cannot ensure the crash consistency of the atomic-write data. Specifically, we consider that the flush for cache line $A$ as shown in Figure~\ref{fig:sequence} is an atomic write, without the data backup based on logging. If a system failure occurs after appending the counter of $A$ (\emph{App(Ac)}) before appending $A$ (\emph{App(A)}) in the write queue, the counter of $A$ (\emph{$A_c$}) is updated and persisted in NVM using the asynchronous DRAM refresh (ADR) mechanism but $A$ is not. After the system is recovered from the failure, the old value of $A$ cannot be decrypted using the new counter. $A$ is an atomic write without data backup, thus resulting in an inconsistent state.

To guarantee the crash consistency of atomic writes in the encrypted NVM, we add a register for each AES circuit. During encrypting the data (i.e., a cache line) evicted from CPU caches, we store its corresponding counter in the register ($Sto(Ac)$) instead of directly appending the counter in the write queue, as shown in Figure~\ref{fig:sequence-with-register}. After finishing encrypting the data, we first store the encrypted data in the register ($Sto(Ac)$), and then simultaneously append the encrypted data and its counter in the write queue ($App(Ac+A)$). We observe that both data and its associated counter simultaneously exist in the write queue or not, by using a register. As a result, the crash consistency of atomic writes is ensured in SecPM. \atc{Moreover, since the size of the register is very small, i.e., 2 cache lines (one for the data cache line and the other for its counter cache line), and reads and writes in such a small register are very fast,} the use of the register has a negligible impact on the system performance.


\subsection{Counter Write Reduction}
\label{design:write-reduction}
\noindent
In the secure NVM, each CPU cache line flush appends two write requests in the write queue, which doubles the number of write requests, compared with an un-encrypted NVM. Thus the performance of the memory system would significantly decrease. SecPM proposes a locality-aware counter write reduction (CWR) scheme to improve the system performance via leveraging the spatial locality of counter storage, log and data writes.

\begin{figure}[t]
  \centering
    \includegraphics [width=0.32\textwidth]{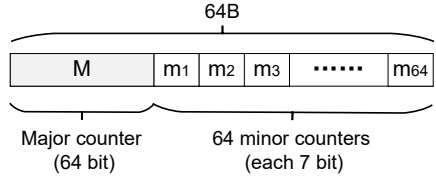}

    \caption{\label{fig:locality-cme} The counter storage of memory lines in a physical page. \textit{(A 4KB page includes 64 memory lines. Each memory line is encrypted using the major counter concatenated with a minor counter.)}}
     \vspace{-7px}
\end{figure}

\subsubsection{Spatial Locality of Counter Storage}
\label{locality_counter_storage}
\noindent
In order to reduce the storage overhead of counters, the counter mode encryption uses a shared major counter (\emph{M}) for an entire page and 64 minor counters ($m_1$, $m_2$, ..., $m_{64}$) each for a memory line in the 4KB page~\cite{awad2016silent,yan2006improving}, as shown in Figure~\ref{fig:locality-cme}. The major counter is 64 bits and each minor counter is 7 bits. Thus the counters of all memory lines in a page are 64B and stored in one memory line, exhibiting good spatial locality.

In a page, each memory line is encrypted by the per-page major counter concatenated with a per-line minor counter. When a memory line is rewritten, its corresponding minor counter increases by one. Although updating a minor counter only modifies several bits, persisting the minor counter has to write the entire memory line into NVM since a memory line is the basic unit of memory writes. When a minor counter overflows, counter mode encryption increases the major counter by one, resets all minor counters to 0, and re-encrypts all memory lines in the page using the new counters~\cite{yan2006improving}. The process of re-encrypting a page is presented in Section~\ref{sec:page-re-encryption}. The 64-bit major counter cannot overflow throughout the lifespan of an NVM, since the count range, i.e., $2^{64} \approx 10^{20}$, is far larger than the cell endurance limit of NVM, e.g., $10^7\sim 10^9$ for PCM~\cite{qureshi2009enhancing,zhou2009durable} and $10^8\sim 10^{12}$ for ReRAM~\cite{lee2010evidence,lee2011fast}.

\subsubsection{Spatial Locality of Log and Data Writes}
\noindent
Since a log is stored in a contiguous region in NVM, the log writes of a transaction flush multiple cache lines which have the contiguous physical addresses, thus having good spatial locality. Moreover, the data writes of a transaction usually have good spatial locality, since programs usually allocate a contiguous memory region for a transaction~\cite{Hudson2006MST}. Hence, the cache lines flushed into the contiguous region have the contiguous physical addresses. For example, a transaction inserts a 1KB key-value item into a key-value store maintained in NVM, which will flush 16 cache lines with the contiguous physical addresses.

\begin{figure}[t]
 \vspace{1px}
    \centering
    \subfloat[The write queue without CWR]{
    \label{fig:write-queue-baseline}
    \includegraphics[width=0.33\textwidth]{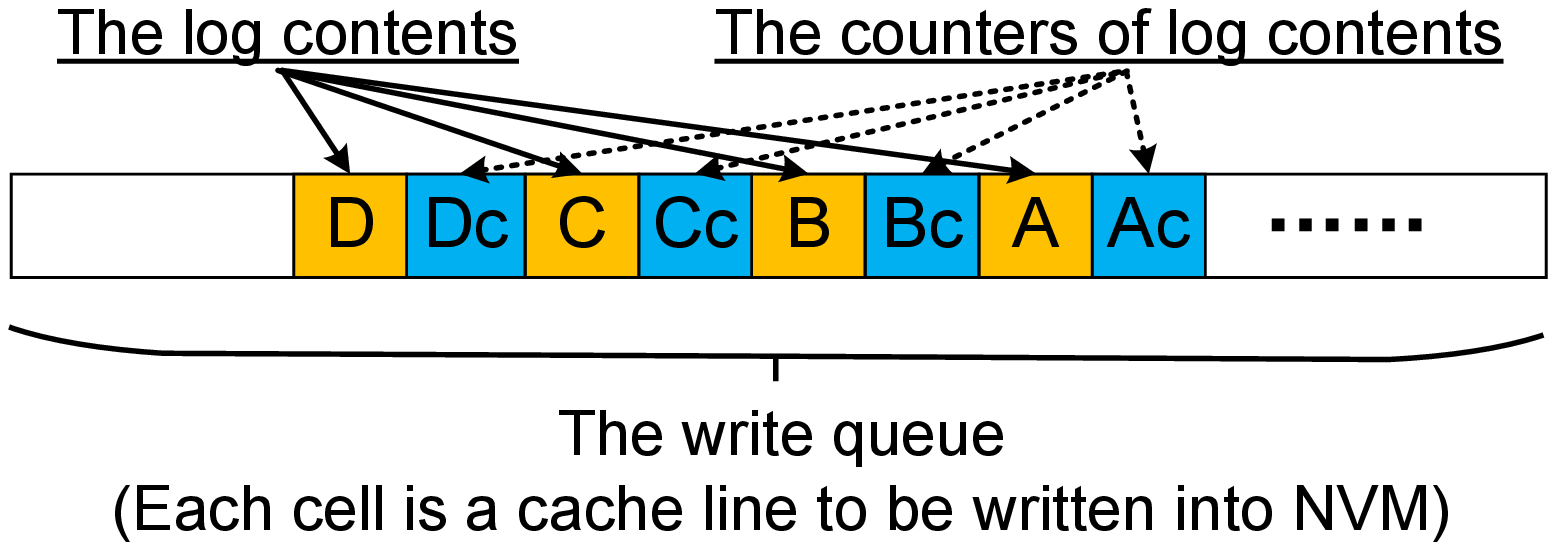}}

    \subfloat[The write queue with CWR]{
    \label{fig:write-queue-CWR}
    \includegraphics[width=0.33\textwidth]{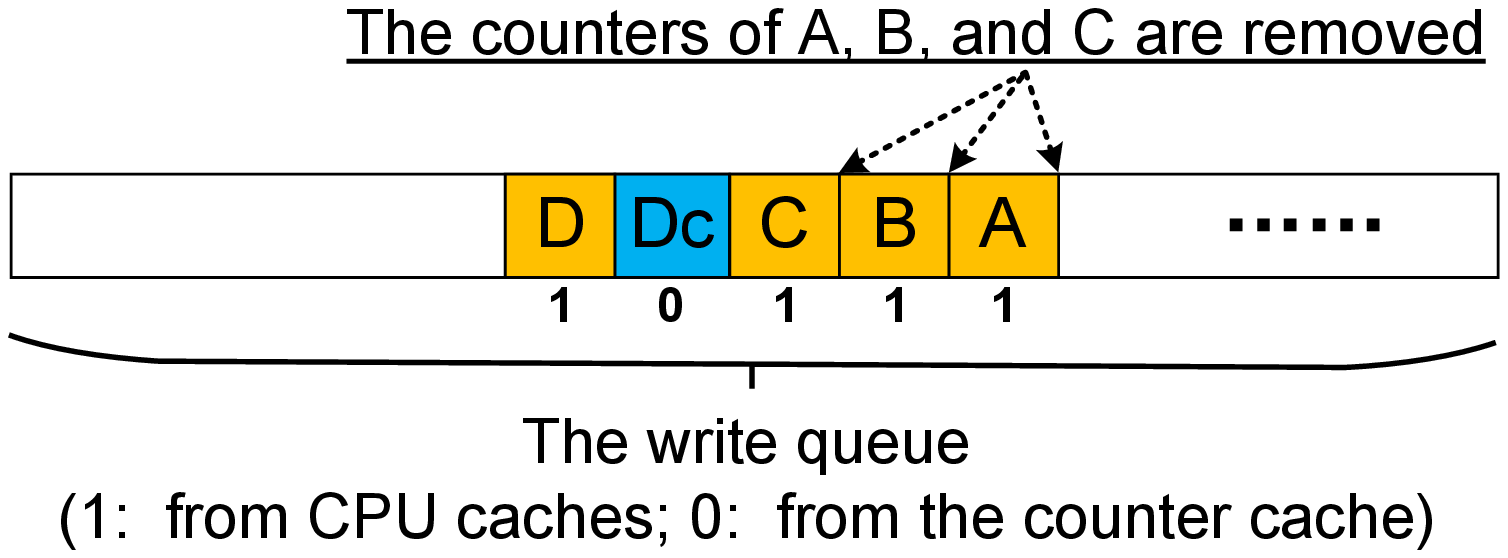}}
    \vspace{-6px}
    \caption{An illustration of the write queue when writing a log with and without CWR scheme.}
    \vspace{-9px}
\end{figure}

\subsubsection{Counter Write Reduction Scheme}
\begin{figure}[t]
    \vspace{4px}
    \centering
    \subfloat[In-order writes]{
    \label{fig:counter-inorder}
    \includegraphics[width=0.23\textwidth]{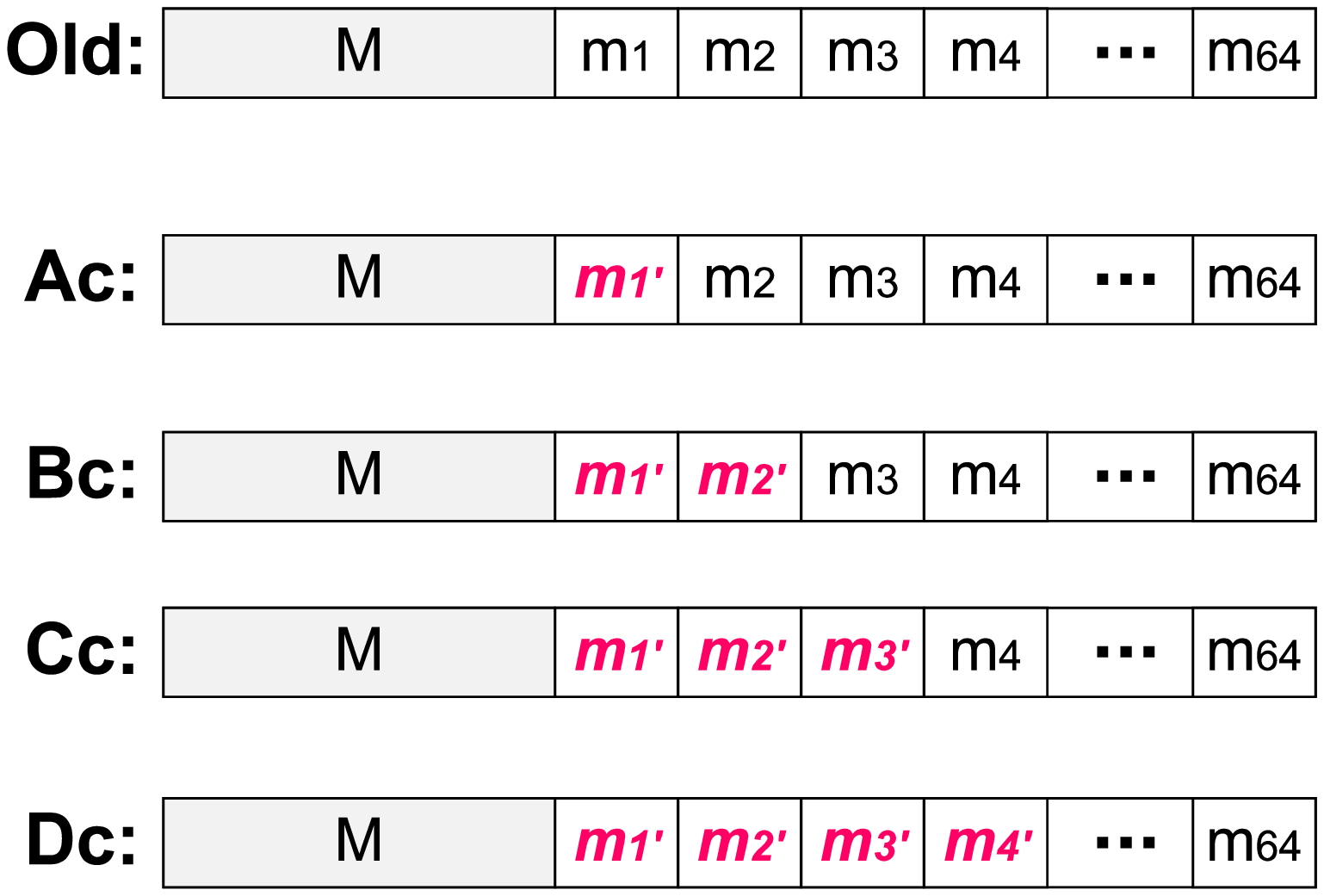}}
    \hspace{0px}
    \subfloat[Out-of-order writes]{
    \label{fig:counter-outorder}
    \includegraphics[width=0.23\textwidth]{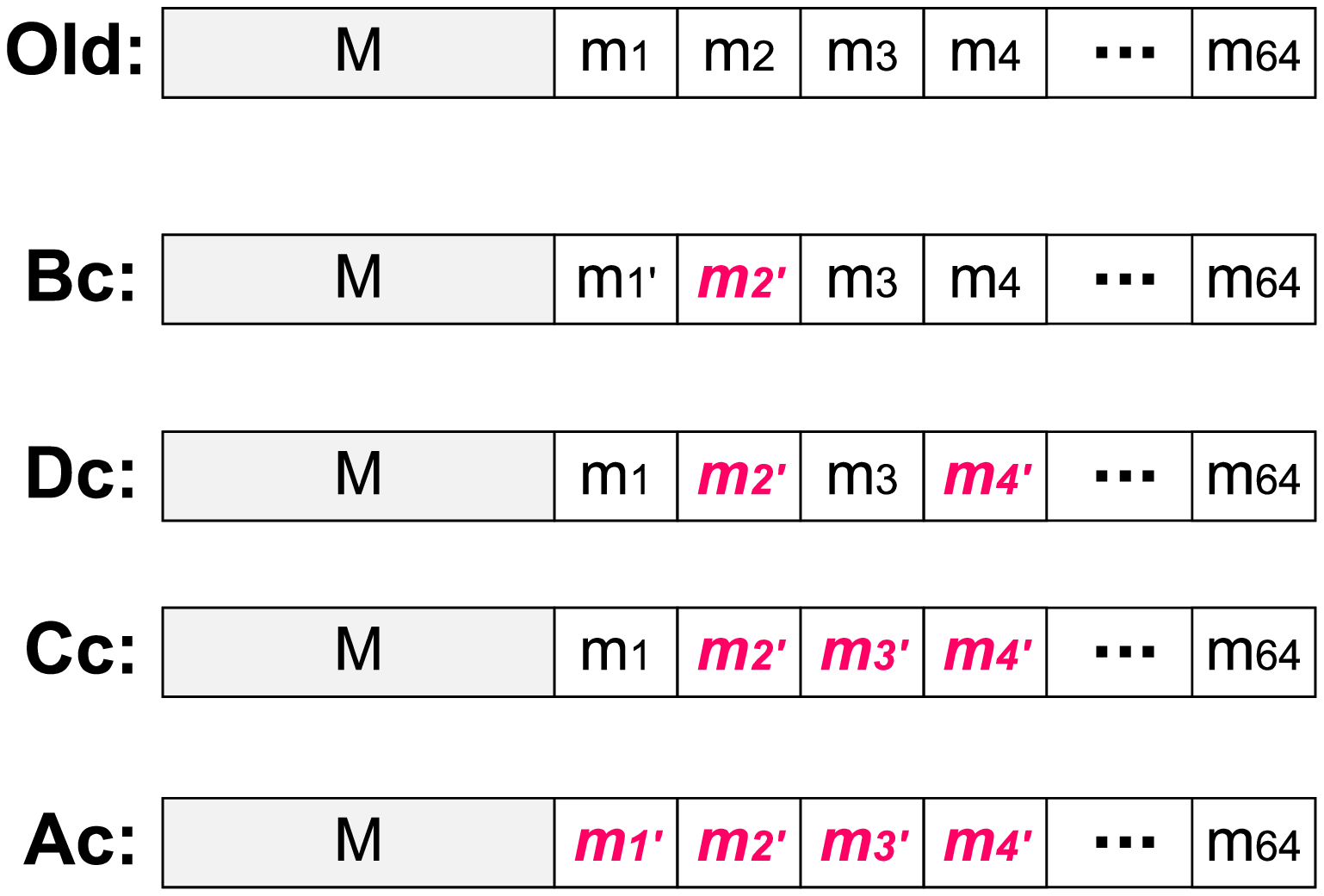}}
     \vspace{-2px}
    \caption{\label{fig:counter-write-contents} The contents of the corresponding counter cache lines when the CPU cache lines \emph{A}, \emph{B}, \emph{C} and \emph{D} are flushed in order or out of order.}
     \vspace{-6px}
\end{figure}

\noindent
Based on the locality of counter storage, log and data writes, we show the write queue during flushing the log entry of a transaction in Figure~\ref{fig:write-queue-baseline}. The log entry contains multiple cache lines (i.e., \emph{A}, \emph{B}, \emph{C}, and \emph{D}) with contiguous physical addresses in the same physical page. Since all counters of a page are contained in one memory line as shown in Figure~\ref{fig:locality-cme}, the counter cache lines of the log entry, i.e., \emph{$A_c$}, \emph{$B_c$}, \emph{$C_c$}, and \emph{$D_c$}, will be written to the same memory line. Moreover, since these counter cache lines are evicted from the write-through counter cache, the latter counter cache lines contain the updated contents of the former ones with the same address.
For example, the memory lines $A$, $B$, $C$ and $D$ correspond to the minor counters $m_1$, $m_2$, $m_3$ and $m_4$, respectively. \emph{$A_c$}, \emph{$B_c$}, \emph{$C_c$}, and \emph{$D_c$} are written into NVM in order, as shown in Figure~\ref{fig:counter-inorder}. We observe that the counter cache line $A_c$ only contains the updated minor counter ${m_1}'$, and the counter cache line $B_c$ contains the updated minor counters ${m_1}'$ and ${m_2}'$, and the counter cache line $C_c$ contains the updated minor counters ${m_1}'$, ${m_2}'$ and ${m_3}'$. Moreover, the multiple cache lines of a log entry, i.e., $A$, $B$, $C$ and $D$, may be flushed from CPU caches out of order. For example, $B$, $D$, $C$ and $A$ are flushed in turn. Thus the corresponding counter cache lines are written into the write queue in the order of $B_c$, $D_c$, $C_c$ and $A_c$. In this case, it is still valid that the counter cache lines written latter contain the updated contents of the former ones, as shown in Figure~\ref{fig:counter-outorder}.

According to above observations and insights, we present a counter write reduction (CWR) scheme. Specifically, when a new counter cache line evicted from the counter cache reaches the write queue, we check whether a counter cache line in the write queue has the same physical address as the new one. If yes, we remove this existing counter cache line without causing any loss of data, since the new counter cache line contains the updated contents of the removed one as shown in Figure~\ref{fig:counter-write-contents}. To reduce the latency of checking the cache lines with the same address, we add a one-bit flag for each cache line in the write queue. The flag is used to distinguish whether a cache line is from CPU caches or the counter cache. , e.g., `1' for the cache lines from CPU caches and `0' for those from the counter cache.
Thus we can check the cache lines only from the counter cache based on the flag. By performing the CWR scheme, the new write queue is shown in Figure~\ref{fig:write-queue-CWR}. We observe that the number of write requests is significantly reduced, since the counters of $A$, $B$ and $C$ are removed.
When a transaction flushes a log with the size of one page, $64*2=128$ CPU and counter cache lines are written into NVM without our proposed CWR scheme.  By using CWR, only $64+1 = 65$ cache lines are written, reducing almost half of NVM writes.

\subsection{Page Re-encryption to Handle Overflow} \label{sec:page-re-encryption}
\noindent
In the counter mode encryption, the major counter cannot overflow, as discussed in Section~\ref{locality_counter_storage}. When the minor counter of a memory line in a page overflows, the page needs to be re-encrypted using the updated major counter. In the following, we first present the page re-encryption process in existing work~\cite{chhabra2011secureme,yan2006improving}, \atc{which however causes the problem of crash inconsistency for persistent memory. We then present how to guarantee crash consistency during page re-encryption in SecPM.}

\reencryption{To re-encrypt a page, all memory lines in this page are read into the last level cache. These memory lines are then re-encrypted one by one using the updated major counter concatenated with a zeroed minor counter, and finally written back into main memory. During re-encrypting these memory lines, a re-encryption status register (RSR) maintained in the memory controller is used to track the re-encryption status of each memory line within a page~\cite{yan2006improving}. The RSR stores the page number and the old major counter of the page. The RSR also maintains a} \verb"done" \reencryption{bit for each memory line within the page to indicate whether the corresponding memory line has already been re-encrypted. After all the 64} \verb"done" \reencryption{bits are set to `1' in the RSR, re-encryption of this page is complete and the RSR is freed.}

Existing work~\cite{yan2006improving} demonstrates that the latency of page re-encryption can be near-completely reduced from the critical path of processor execution, thus having a negligible impact on the system performance. Since the RSR tracks the re-encryption status of each line within a page that is being re-encrypted, the CPU caches are able to continue to service regular memory access \atc{(read and flush) requests to other pages} and the processor is not stalled. For an access to a memory line in the page that is being re-encrypted, there are two cases based on the re-encryption status of the line recorded in the RSR. 1) If the line has been already re-encrypted, i.e., its \verb"done" bit is `1', the access normally proceeds. 2) If the line is not re-encrypted, i.e., its \verb"done" bit is `0', the access simply waits for the re-encryption of the line. Moreover, Huang et al.~\cite{huang2017wear} and \atc{Swami et al.~\cite{swami2018acme} show that the counter mode encryption can be combined with NVM wear-level techniques~\cite{qureshi2009enhancing,Seong2010SRP}, and thus the frequency of page re-encryption is significantly reduced.}

\reencryption{However, when the page re-encryption process is executed in persistent memory, if a system failure occurs during re-encrypting a page, some memory lines within the page have been re-encrypted but others have not. Moreover, the re-encryption status and page number recorded in the RSR are lost. After recovery, the system does not know which page is being re-encrypted and which memory lines in this page have not been re-encrypted. As a result, the memory lines that have not been re-encrypted cannot be correctly decrypted, thus resulting in an inconsistent state.}

\begin{figure*}[t]
    \centering
    \subfloat[64B transaction size]{
    \label{fig:evaluation-write-reduction-64B}
    \includegraphics[width=0.4\textwidth]{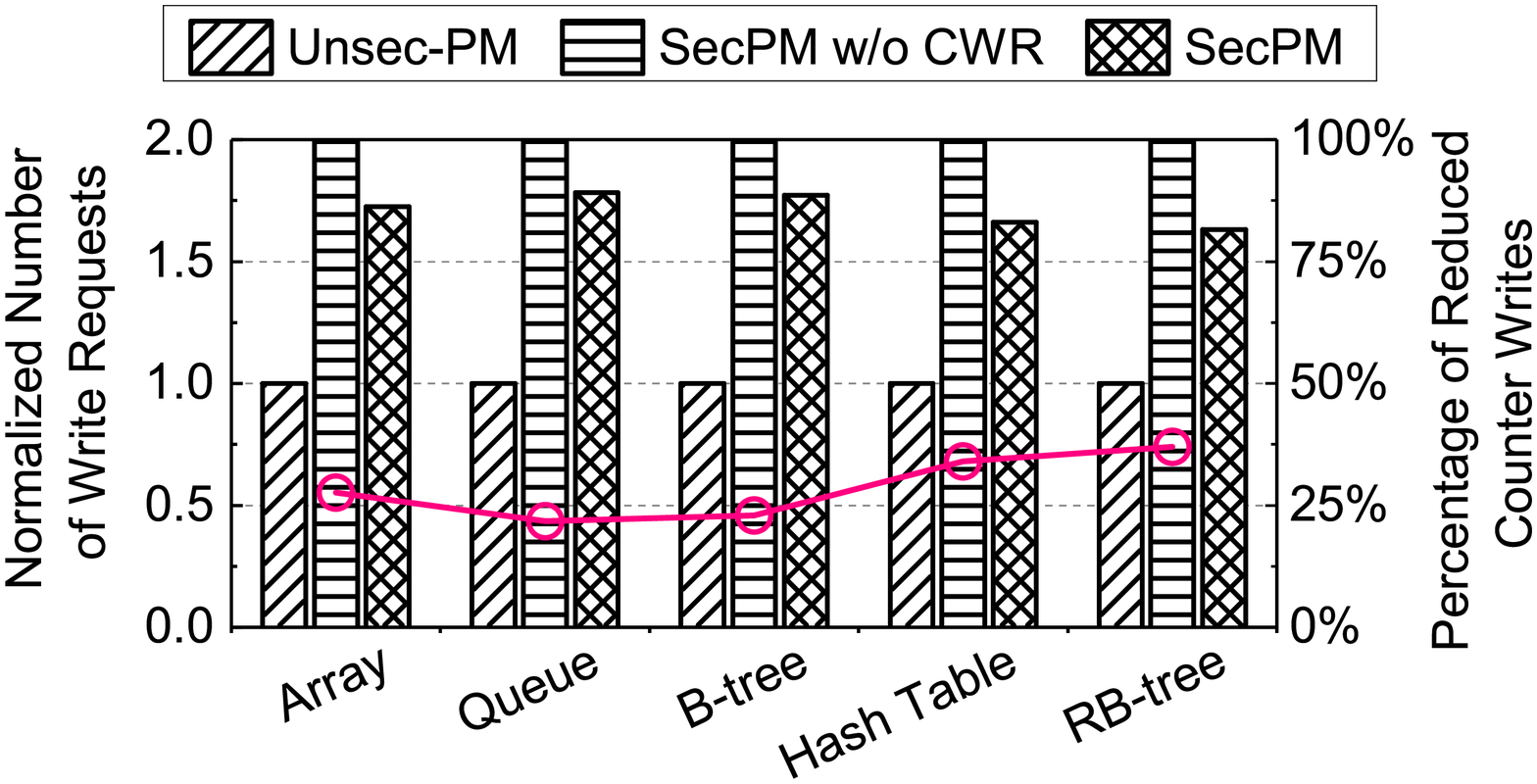}}
    \hspace{45px}
    \subfloat[256B transaction size]{
    \label{fig:evaluation-write-reduction-256B}
    \includegraphics[width=0.4\textwidth]{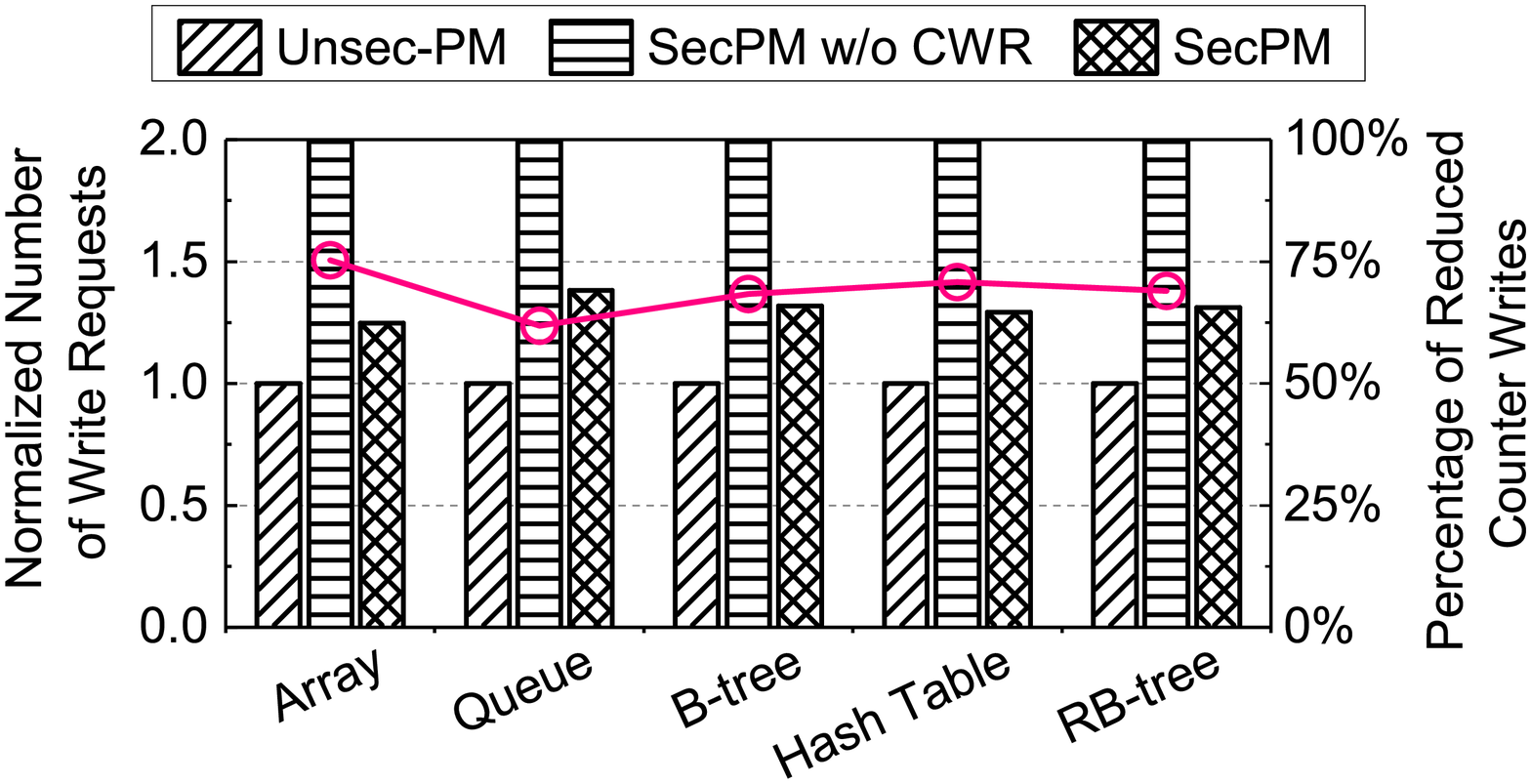}}

    \subfloat[1KB transaction size]{
    \label{fig:evaluation-write-reduction-1KB}
    \includegraphics[width=0.4\textwidth]{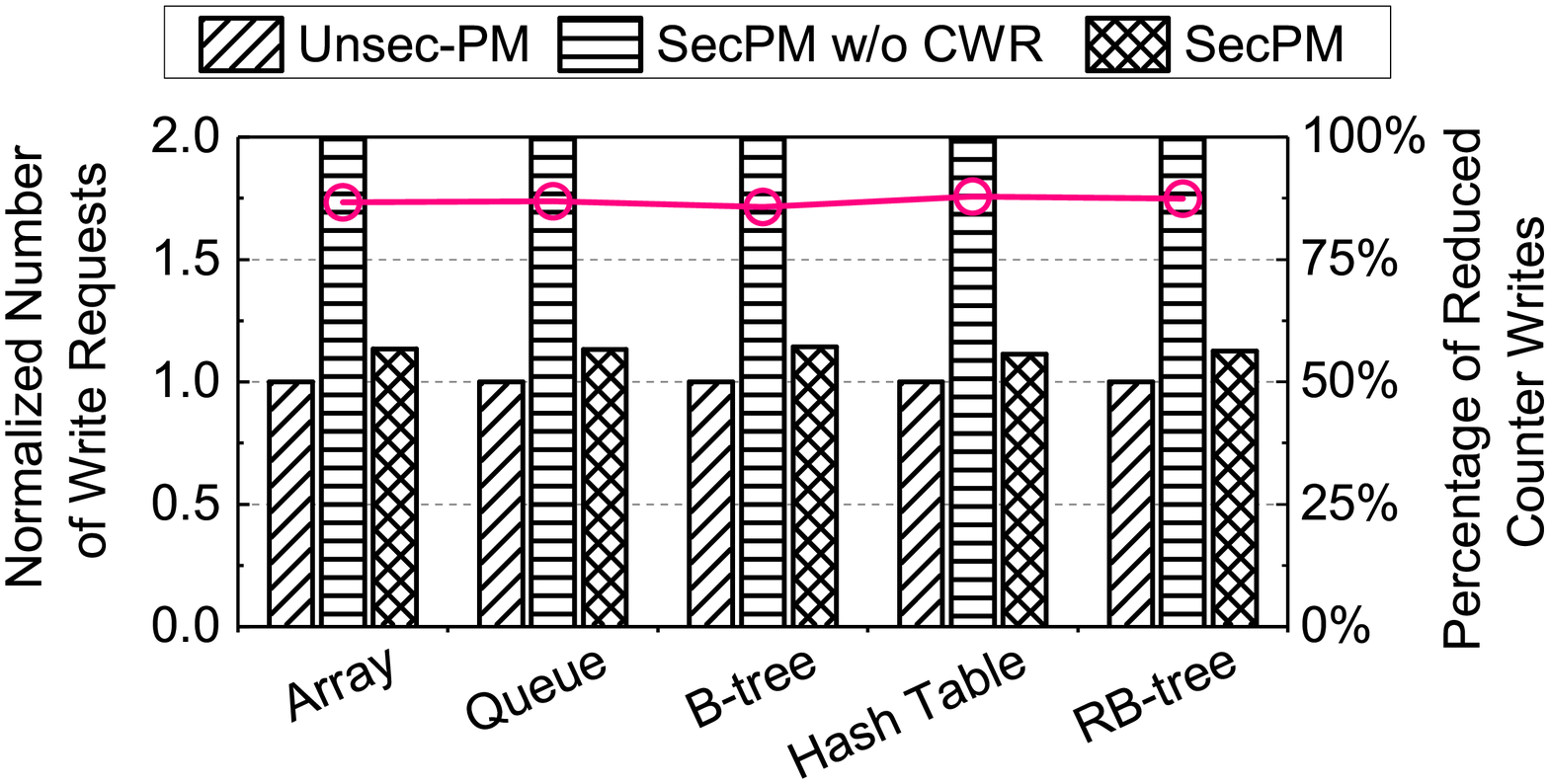}}
    \hspace{45px}
    \subfloat[4KB transaction size]{
    \label{fig:evaluation-write-reduction-4KB}
    \includegraphics[width=0.4\textwidth]{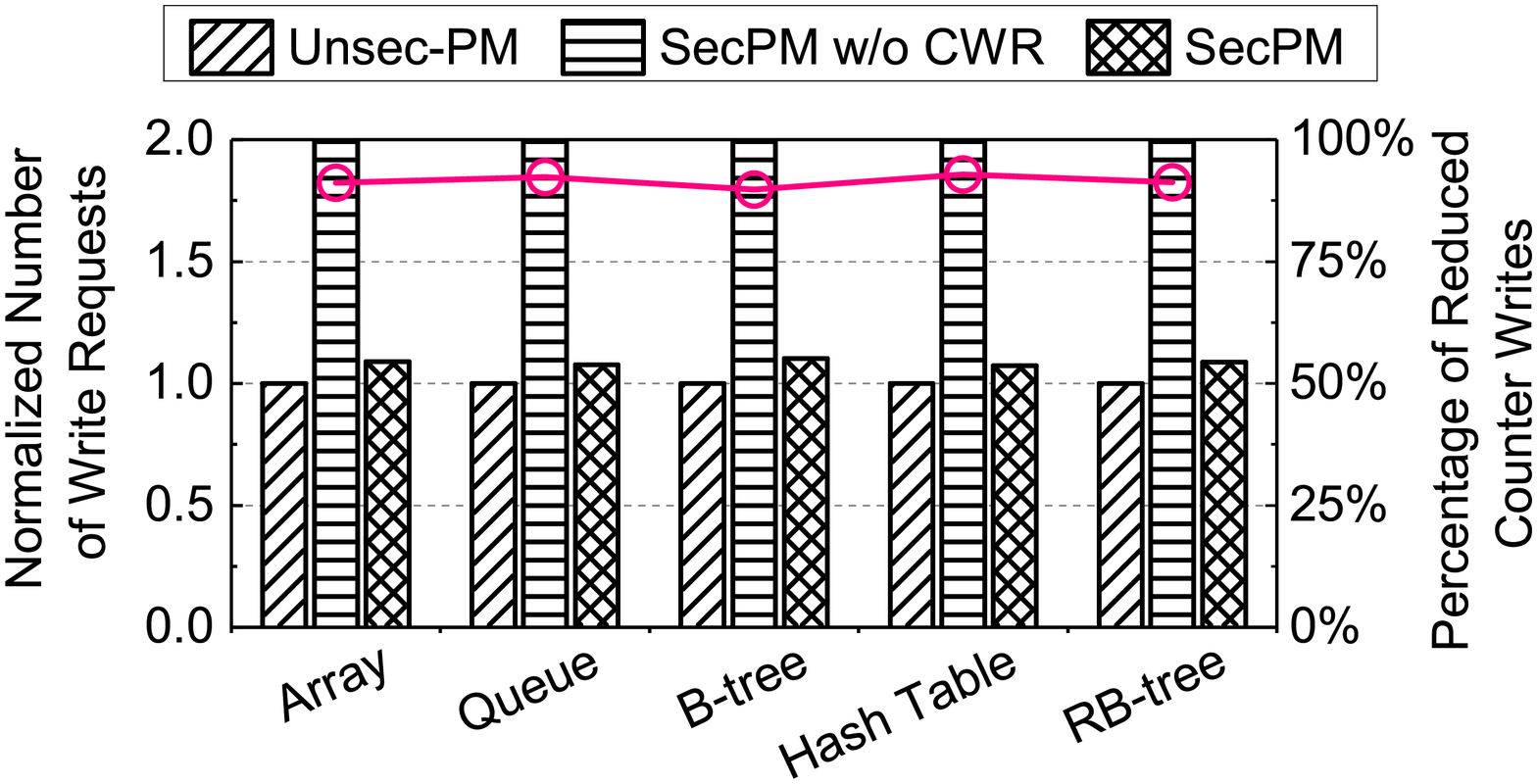}}
    \vspace{-5px}
    \caption{\label{fig:evaluation-write-reduction} The numbers of NVM write requests normalized to those of Unsec-PM with different transaction sizes. \emph{(The black line with circles in each figure shows the percentage of reduced counter writes via CWR in SecPM.)}}
 \vspace{-8px}
\end{figure*}

\reencryption{To guarantee crash consistency of page re-encryption, we employ the ADR mechanism (providing battery backup for the write queue) on the RSR. The battery overhead is negligible, since the size of RSR is very small, i.e., 20 bytes, including 32-bit physical page number, 64-bit old major counter and 64-bit} \verb"done" \reencryption{bits. The data stored in the RSR are flushed into NVM in case of a system failure using the ADR and loaded into RSR again when the system is recovered. Thus the system knows which page is being re-encrypted and which memory lines in the page have not been re-encrypted, based on the content of the RSR. Hence, the system can continue to complete the page re-encryption after the recovery from a system failure. Moreover, the process of re-encrypting each memory line is the same as that the memory controller deals with a regular cache line flush as shown in Figure~\ref{fig:sequence-with-register}. Thus the consistency of writing each re-encrypted line is also ensured by the CWT scheme, and the performance of page re-encryption is also improved by the CWR scheme, since the writes during a page re-encryption have good spatial locality.}






\section{Performance Evaluation}
\label{section4}


\subsection{Methodology}
\noindent
\atc{As real hardware is not available yet for implementing NVMM and the proposed persistence and encryption schemes, we use gem5~\cite{binkert2011gem5} with NVMain~\cite{poremba2015nvmain} to evaluate SecPM.} NVMain is a cycle-accurate main memory simulator for emerging NVM technologies. The NVM system consists of x86-64 processors running at 2GHz, 32KB L1 data and instruction caches, 512KB L2 caches, and 4MB L3 cache. The counter cache is 1MB.  Without loss of generality, we model PCM technologies~\cite{choi201220nm} with 16GB capacity. The PCM latency model is the same as that used in Xu el al.'s work~\cite{xu2015overcoming}. The encryption and decryption latencies of AES circuit are 40ns~\cite{liucrash2018,shi2005high}. To support the simulation of persistent memory, we employ the \verb"clwb" and \verb"sfence" instructions that have been implemented in the latest gem5.

\begin{table}[t]
\vspace{5px}
\footnotesize
\caption{\label{table:configure} The configurations of the NVM system.}
\vspace{-5px}
\begin{center}
\begin{tabular}{|c|l|}
    \hline
     \multicolumn{2}{|c|}{\textbf{Processor}} \\
    \hline
    CPU & 4 cores, X86-64 processor, 2 GHz        \\
    \hline
    Private L1  cache & 64KB, 8-way, LRU, 2-CPU-cycle latency              \\
    \hline
    Private L2 cache & 512KB, 8-way, LRU, 8-CPU-cycle latency        \\
    \hline
    Shared L3 cache & 4MB, 8-way, LRU, 30-CPU-cycle latency            \\
    \hline

    \multicolumn{2}{|c|}{\textbf{Main Memory}} \\
    \hline
     Capacity &  16GB, 16 banks in 2 ranks     \\
    \hline
    PCM latency model &   \makecell[tl]{tRCD/tCL/tCWD/tFAW/tWTR/tWR  \\
        =48/15/13/50/7.5/300 ns }     \\
    \hline
    En/decryption latency & 40 ns           \\
    \hline
    Write queue & 32 entries                \\
    \hline
    Counter cache & 1 MB, 8-way, LRU, 12-CPU-cycle latency\\
    \hline
\end{tabular}
\end{center}
 \vspace{-7px}
\end{table}

Since the performance improvement of SecPM mainly comes from the counter write reduction (CWR) scheme as presented in Section~\ref{design:write-reduction}, we compare \verb"SecPM" with the \verb"SecPM w/o CWR" which indicates the SecPM without the proposed CWR scheme.
Moreover, we also consider an unsecure persistent memory without memory encryption (\verb"Unsec-PM") as a baseline for comparisons. We do not compare the performance of programmer-transparent SecPM with Liu et al.'s work~\cite{liucrash2018} due to no open-source codes and different design goals.

\begin{figure*}[t]
    \centering
    \subfloat[64B transaction size]{
    \label{fig:evaluation-latency-64B}
    \includegraphics[width=0.38\textwidth]{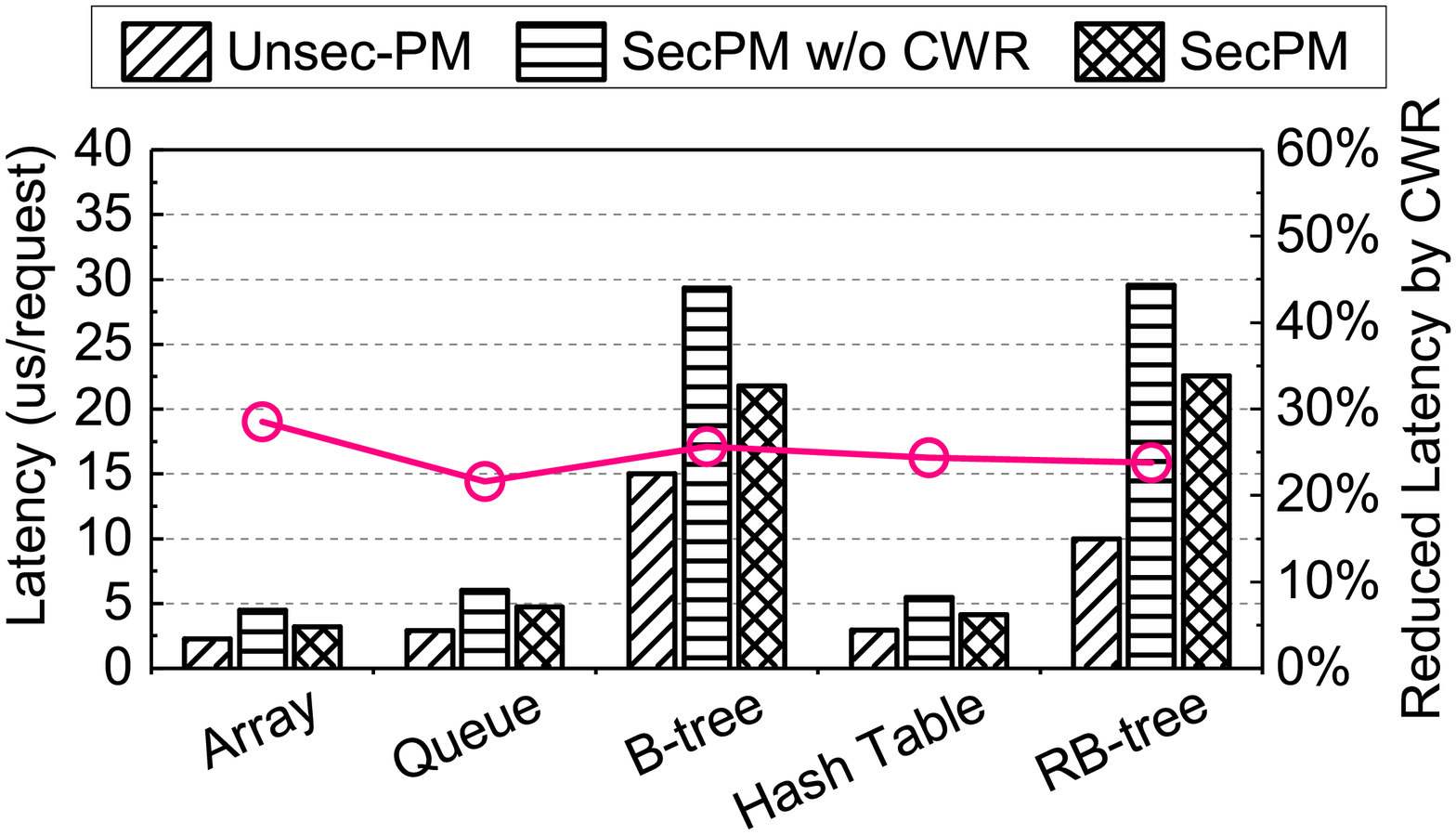}}
    \hspace{45px}
    \subfloat[256B transaction size]{
    \label{fig:evaluation-latency-256B}
    \includegraphics[width=0.38\textwidth]{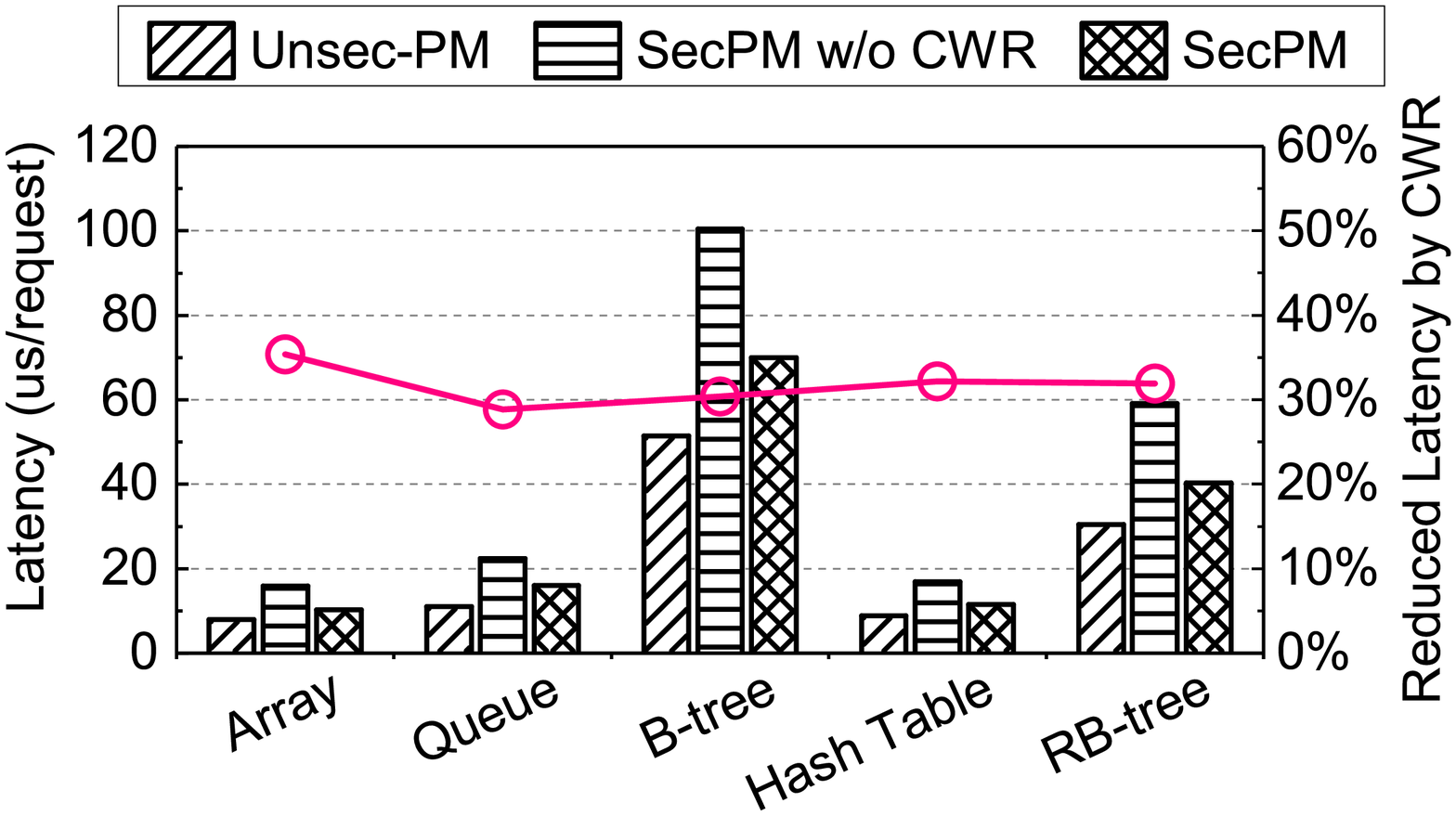}}

    \subfloat[1KB transaction size]{
    \label{fig:evaluation-latency-1KB}
    \includegraphics[width=0.38\textwidth]{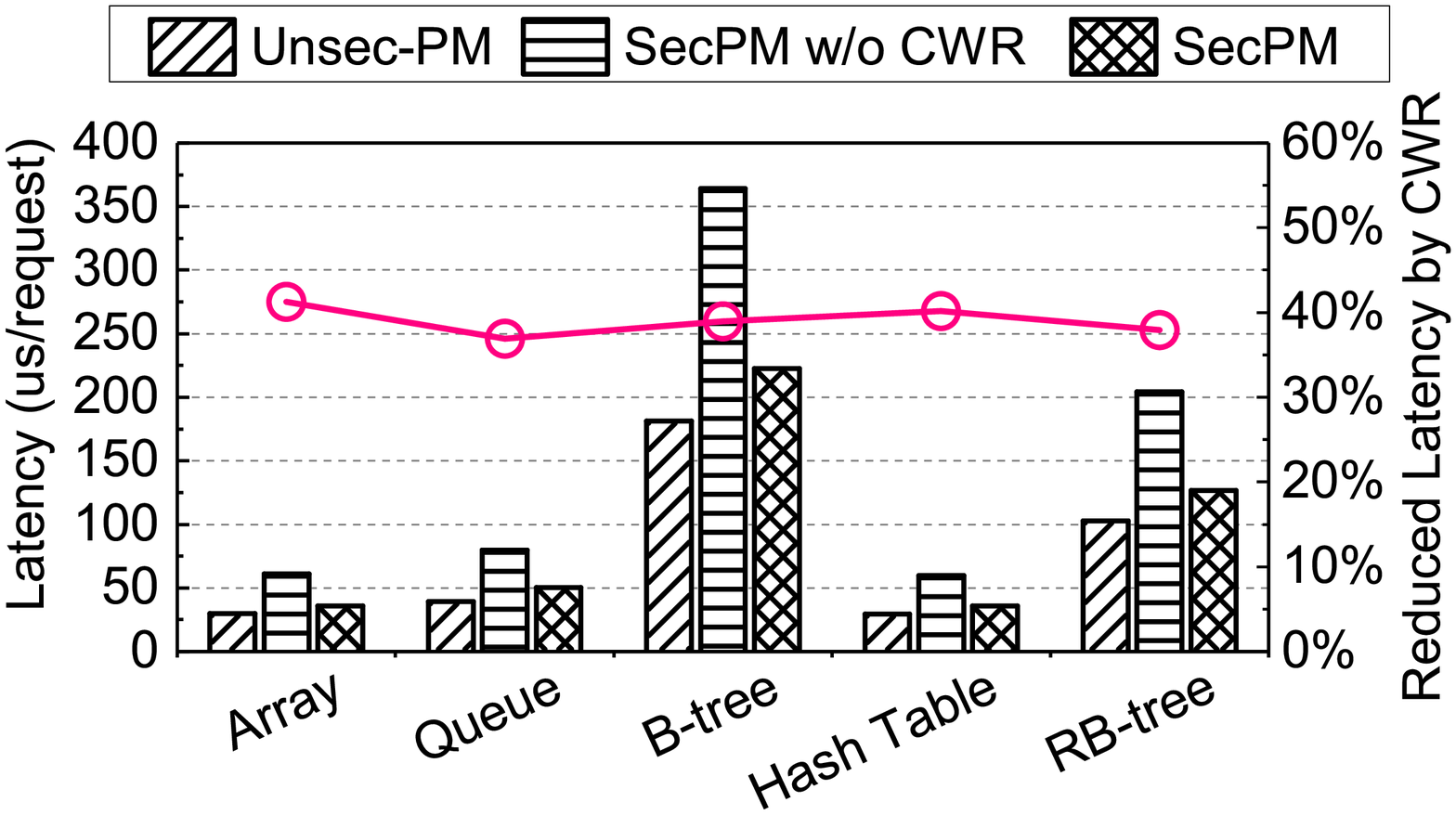}}
    \hspace{45px}
    \subfloat[4KB transaction size]{
    \label{fig:evaluation-latency-4KB}
    \includegraphics[width=0.38\textwidth]{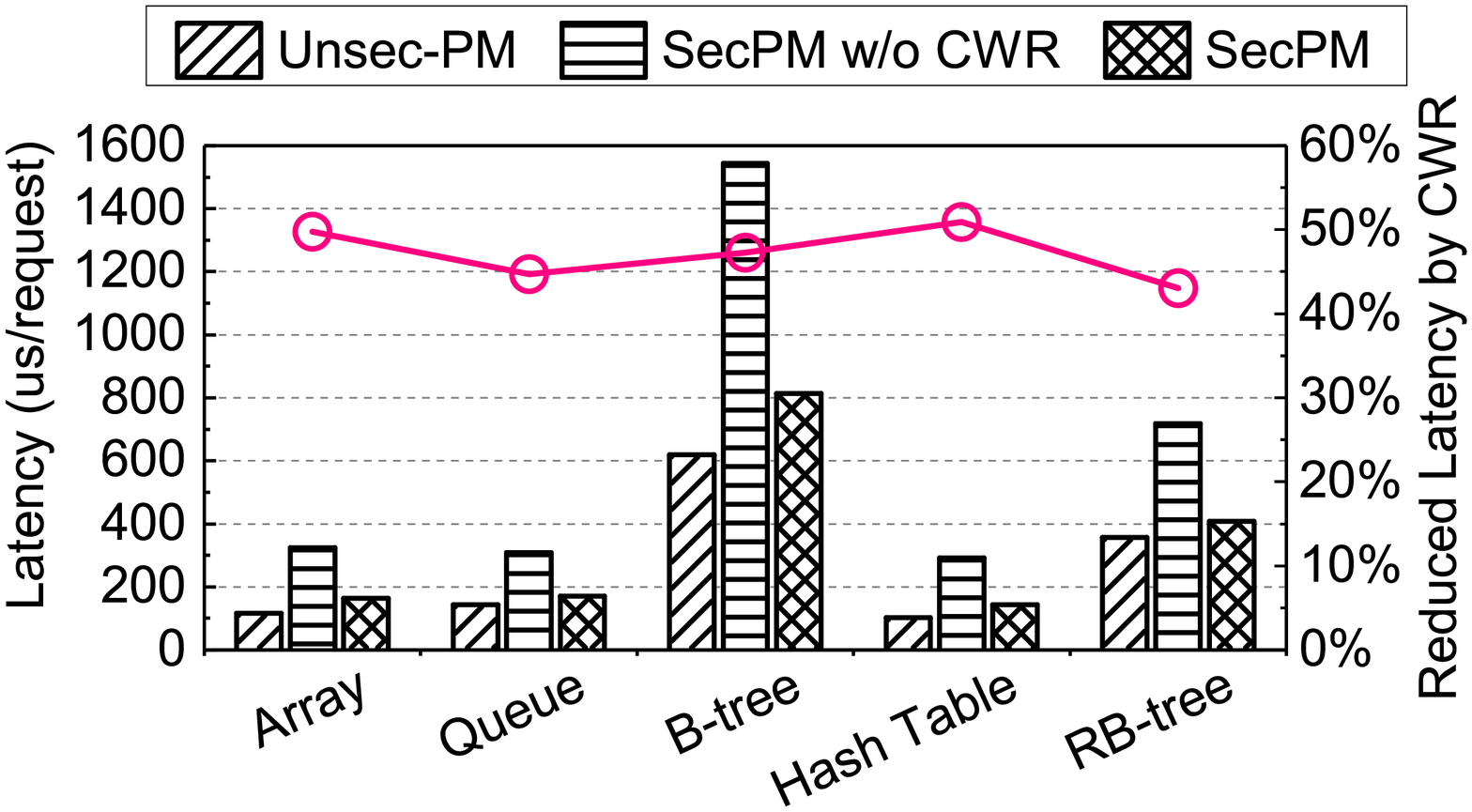}}

    \caption{\label{fig:evaluation-latency} The average latency of executing each transaction with different transaction sizes. \emph{(The black line with circles in each figure shows the percentage of reduced latency that SecPM uses CWR)}.}
 \vspace{-8px}
\end{figure*}

We use five workloads to evaluate the performance of SecPM as listed in the following. The five workloads are also widely used in existing work on persistent memory~\cite{coburn2011nv,kolli2017language,kolli2016delegated,liucrash2018,ren2015thynvm}.
\begin{itemize}
\vspace{-2px}
  \item \textbf{\emph{Array.}} Initializing a 1GB array and then randomly swapping entries.
  \vspace{-2px}
  \item \textbf{\emph{Queue.}} Randomly enqueueing and dequeueing entries in a 1GB queue.
  \vspace{-2px}
  \item \textbf{\emph{B-tree.}} Inserting random key-value items into a 2GB B-tree based key-value store.
  \vspace{-2px}
  \item \textbf{\emph{Hash table.}} Inserting random key-value items into a 2GB hash table based key-value store.
  \vspace{-2px}
  \item \textbf{\emph{RB-tree.}} Inserting random key-value items into a 2GB red-black tree.
\end{itemize}
The ACID property (atomicity, consistency, isolation, and durability) of operations in these workloads is ensured via durable transaction, like existing work~\cite{liu2017dudetm,liucrash2018,nalli2017analysis,ren2015thynvm}.

\subsection{Experimental Results}
\noindent
We first present the experimental results of Unsec-PM, SecPM, and SecPM w/o CWR  in terms of the the number of write requests and transaction execution latency. We then evaluate the sensitivity of the experimental results under different configuration parameters.

\subsubsection{The Number of Write Requests}
\label{evaluation:num_writes}
\noindent
Since the proposed CWR scheme improves the system performance by employing the spatial locality of log and data writes, different transaction request sizes exhibit different spatial locality, hence having a great impact on the system performance. Thus we vary the transaction request sizes in the five workloads from 64B to 4KB to evaluate the number of NVM writes.

Figure~\ref{fig:evaluation-write-reduction} shows the number of write requests to NVM normalized to that of Unsec-PM in the five workloads. We observe the SecPM w/o CWR achieves the security and crash consistency but incurs about $2 \times$ writes compared with Unsec-PM, whatever the transaction request size is. The reason is that each data write in secure NVM produces two write requests: one for the data and the other for the counter.
Compared with the SecPM w/o CWR, SecPM significantly reduces the number of NVM writes. Even in the worst case where the request size is 64B (which means only a cache line is written in a transaction), SecPM reduces $22\%-37\%$ of counter writes compared with the SecPM w/o CWR, since the transaction data writes have no locality while the log writes have locality.
When the transaction request size increases, the locality of data writes significantly increases. SecPM reduces $62\%-75\%$, $86\%-88\%$, and $90\%-93\%$ of counter writes compared with the SecPM w/o CWR, when the transaction sizes are 256B, 1KB, and 4KB, respectively.

\subsubsection{Transaction Execution Latency}
\noindent
We vary the transaction request sizes in the five workloads from 64B to 4KB to evaluate the transaction execution latency.

Figure~\ref{fig:evaluation-latency} shows the average latency of executing each transaction with different transaction sizes. We observe that the SecPM w/o CWR increases the transaction execution latency by $2-3 \times$ across these workloads compared with Unsec-PM, since doubling the number of write requests to NVM significantly degrades the system performance. Compared with the SecPM w/o CWR, SecPM significantly reduces the average transaction execution latency via reducing the number of counter writes. In the worst case where the transaction size is 64B, SecPM still reduces the transaction execution latency by $20\%-29\%$ compared with the SecPM w/o CWR, due to reducing $22\%-37\%$ of counter writes as evaluated in Section~\ref{evaluation:num_writes}. When the transaction sizes are 256B, 1KB, and 4KB, SecPM reduces $29\%-36\%$, $37\%-41\%$, and $43\%-51\%$ of average execution latency respectively, compared with the SecPM w/o CWR.
When the transaction sizes are larger than 1KB, SecPM incurs only a little latency increase, compared with Unsec-PM, which comes from the the latency overhead of data encryption and decryption. In summary, SecPM speeds up the transaction execution by $1.3\times-2.0\times$ via the CWR scheme, due to the reduction in the number of write requests, \atc{and achieves the performance close to an unsecure persistent memory system for large transactions.}

\subsubsection{Sensitivity to Write Queue Size}
\noindent
We use the fixed configurations of 1MB counter cache and 1KB transaction size, and vary the write queue length from 8 to 128 to evaluate the performance in terms of the number of write requests and transaction execution latency.

\begin{figure}[t]
    \centering
    \subfloat[The percentage of reduced counter writes]{
    \label{fig:evaluation-write_queue_counter_writes}
    \includegraphics[width=0.43\textwidth]{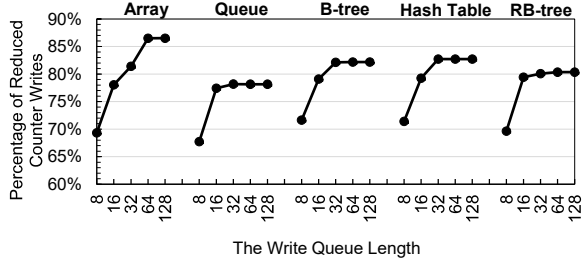}}

    \subfloat[The average latency of executing each transaction]{
    \label{fig:evaluation-write_queue_latency}
    \includegraphics[width=0.43\textwidth]{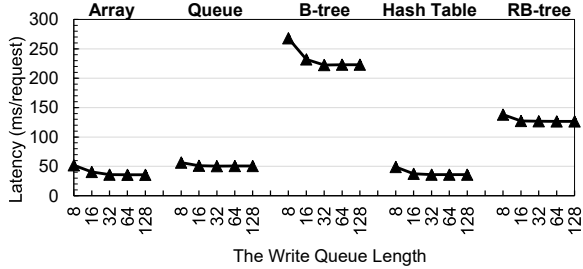}}

    \caption{\label{fig:evaluation-write_queue} The  percentage of reduced counter writes and average latency of executing each transaction with different transaction sizes.}
     \vspace{-8px}
\end{figure}

\begin{figure}[t]
    \centering
    \subfloat[Counter cache hit rate]{
    \label{fig:evaluation-counter-cache-hit-rate}
    \includegraphics[width=0.43\textwidth]{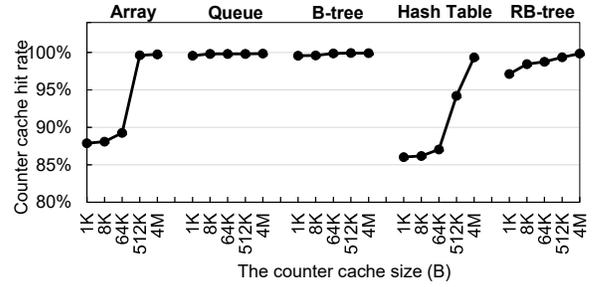}}

    \subfloat[Speedup over a 1KB counter cache]{
    \label{fig:evaluation-counter-cache-speedup}
    \includegraphics[width=0.43\textwidth]{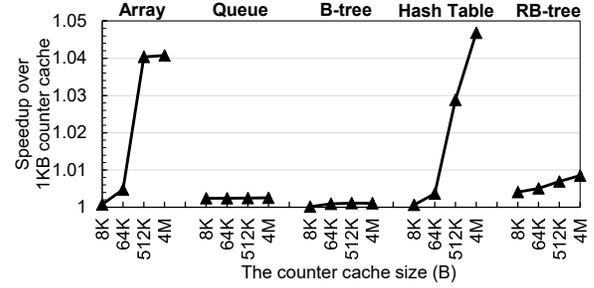}}

    \caption{\label{fig:evaluation-write_queue} Evaluating SecPM with different counter cache sizes.}
    \vspace{-8px}
\end{figure}

Figure~\ref{fig:evaluation-write_queue_counter_writes} shows the influence of different write queue lengths on the percentage of reduced counter writes in SecPM, compared with the SecPM w/o CWR. We observe that SecPM reduces more counter writes with longer write queue. The reason is that longer write queue provides more opportunities for the CWR scheme to find and merge more counter writes with the same physical address in the write queue. For most workloads including queue, B-tree, hash table and RB-tree, when the write queue length is larger than 32, the percentage of reduced counter writes increases little. When the write queue length increases from 8 to 128, SecPM reduces $17\%$, $11\%$, $11\%$, $11\%$, and $10\%$ of more counter writes for array, queue, B-tree, hash table, and RB-tree workloads, respectively.

Figure~\ref{fig:evaluation-write_queue_latency} shows the influence of different write queue lengths on the transaction execution latency in SecPM. We observe that increasing the write queue length decreases the average latency of executing each transaction in all workloads, since longer write queue reduces more counter writes. When the write queue length increases from 8 to 128, SecPM reduces the average transaction execution latency by $31\%$, $11\%$, $17\%$, $27\%$, and $9\%$ for array, queue, B-tree, hash table, and RB-tree workloads, respectively.

\subsubsection{Sensitivity to Counter Cache Size}
\noindent
We use the fixed configurations of 32-entry write queue and 1KB transaction size, and vary the counter cache size from 1KB to 4MB to evaluate the counter cache hit rate and workload execution time.

Figure~\ref{fig:evaluation-counter-cache-hit-rate} shows the influence of different counter cache sizes on the cache hit rate in SecPM. We observe that increasing the counter cache size has a great impact on the counter cache hit rates for array, hash table, and RB-tree, but rarely affects those of queue and B-tree. The reason is that the dequeue or enqueue in the queue accesses a continuous memory space and B-tree has the structure that a node continuously stores multiple key-value items, which exhibit good spatial locality for data accesses. In contrast, the random entry swaps in the array, inserting items into random hash locations in the hash table, and the structure of one item per node in the RB-tree exhibit poor spatial locality for data accesses. The good spatial locality for data accesses produces high counter cache hit rate. When reading a memory line in a page, all counters that decrypt this page are loaded into the counter cache, due to being stored in a memory line. The following accesses to the same page always hit the counter cache. As shown in Figure~\ref{fig:evaluation-counter-cache-hit-rate}, when the counter cache size increases from 1KB to 4MB, the cache hit rate is improved by $12\%$, $14\%$, and $3\%$ for array, hash table and rb-tree workloads.

Figure~\ref{fig:evaluation-counter-cache-speedup} shows the influence of different counter cache sizes on the overall execution time of workloads in SecPM. The execution time of workloads under different counter cache sizes are normalized to those under 1KB counter cache size. We observe that different counter cache sizes have little impact on the execution time of queue and B-tree workloads. For array, hash table, and RB-tree, the execution performance is improved respectively by $4\%$, $5\%$, and $1\%$, when the counter cache size increases from 1KB to 4MB.


\subsubsection{Multiple-core Performance}
\begin{figure}[t]
    \vspace{5px}
  \centering
    \includegraphics [width=0.44\textwidth]{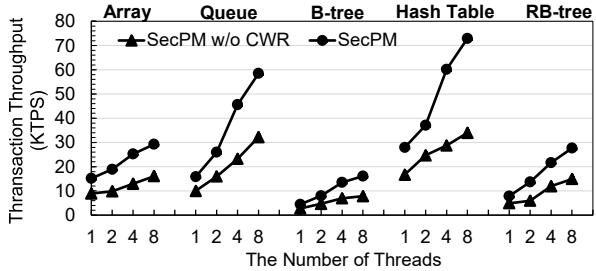}
    \caption{\label{fig:evaluation-concurrency} The concurrent transaction throughput in the 1/2/4/8-core systems.}
\end{figure}

\noindent
We evaluate the performance of SecPM in a multi-core system, where each
thread executes the same transactions on different cores. We use the configurations of 32-entry write queue, 1KB transaction size, and 1MB counter cache.
Figure~\ref{fig:evaluation-concurrency} shows the transaction throughput of different workloads in the 1/2/4/8-core systems. We observe that the transaction throughput of workloads in SecPM scales well with the increase of the number of cores. Moreover, compared with the SecPM w/o CWR, SecPM improves the transaction throughput by $64\%$, $81\%$, $90\%$ and $95\%$ on average in a 1/2/4/8-core system respectively.


\vspace{2px}
\section{Related Work}
\vspace{2px}
\label{section5}
\noindent


\emph{\textbf{$\bullet$ Secure Non-volatile Memory.}} As NVM suffers from the data remanence vulnerability, data security in NVM has been widely studied. DEUCE~\cite{young2015deuce} proposes a dual-counter encryption scheme to reduce the write traffic in the encrypted NVM by re-encrypting only the modified words in a memory line. Based on DEUCE, SECRET~\cite{swami2016secret} further avoids the re-encryption of zero-content words in a memory line to reduce bit writes. Silent Shredder~\cite{awad2016silent} reduces NVM writes in the encrpted NVM by eliminating the full-zero cache line writes produced from data shredding. DeWrite~\cite{zuo2018dewrite} proposes a lightweight deduplication scheme to enhance the performance and endurance of the encrypted NVM via eliminating duplicate-content writes.
All these schemes on the encrypted NVM mainly focus on reducing the writes of encrypted data to NVM, which do not consider the crash consistency in the secure NVM. \atc{Moreover, some existing works focusing on memory authentication in NVM~\cite{rakshit2017assure,ye2018osiris}, which are orthogonal to our work, as discussed in Section~\ref{threat-model}.}

\emph{\textbf{$\bullet$ Crash Consistency in Non-volatile Memory.}} To achieve data persistence, various durable transaction systems, such as NV-Heaps~\cite{coburn2011nv}, Mnemosyne~\cite{volos2011mnemosyne}, DudeTM~\cite{liu2017dudetm}, NVML~\cite{nvml2018}, and DCT~\cite{kolli2016high}, are proposed to manage persistent data with crash consistency in NVM.
Moreover, multiple NVM-based file systems, such as BPFS~\cite{condit2009better}, PMFS~\cite{dulloor2014system}, Mojim~\cite{zhang2015mojim}, NOVA~\cite{xu2016nova}, and NOVA-Fortis~\cite{xu2017nova}, are proposed to achieve the improvement of storage performance by leveraging the byte-addressable benefit of NVM, which also provide the crash consistency guarantee by employing copy-based techniques, e.g., logging, copy-on-write (shadowing page), and replication.
All these schemes are built on the un-encrypted NVM, without considering the memory encryption on NVM.

SecPM aims to ensure both the security and crash consistency of NVM. In terms of addressing the crash consistency of secure NVM, the most related work comes from Liu et al.~\cite{liucrash2018}, which \atc{is the first work to efficiently ensure that data and its counter are atomically persisted via the selective counter-atomicity.}
For a different design goal, i.e., programmer transparency, SecPM performs only slight modifications on the memory controller to achieve crash consistency. \atc{Moreover, SecPM achieves significant performance improvement and write reduction via a counter write reduction scheme.}

\section{Conclusion}
\label{section6}
\noindent
This paper proposes SecPM to achieve both the security and persistence in non-volatile main memory. SecPM leverages a counter cache write-through scheme with a register to guarantee crash consistency of durable transaction as well as atomic writes in secure NVM. Moreover, a counter write reduction (CWR) scheme is introduced to improve the system performance. These schemes are implemented with slight modifications only on the hardware layer, which are transparent for programmers and applications. \atc{Thus programs and applications running on an un-encrypted NVM can be directly executed on a secure NVM with SecPM.} Experimental results demonstrate that SecPM \atc{achieves the performance close to an unsecure persistent memory system for large transactions.}

{
\bibliographystyle{acm}
\bibliography{references}}


\end{document}